\documentclass[12pt]{article}
\usepackage{amssymb}

\def\be{\begin{equation}}
\def\ee{\end{equation}}
\def\bea{\begin{eqnarray}}
\def\eea{\end{eqnarray}}

\def\mod{{\,\,\rm mod}\,\,}
\def\L{{\rm L}}
\def\N{{\rm N}}
\def\K{{\rm K}}
\def\Z{{\mathbb{Z}}}
\def\calN{{\mathcal N}}

\def\M{{\rm M}}

\def\k{{\bf k}}
\def\mod{{\,\,\rm mod}\,\,}

\usepackage{epsfig,multicol,bbm}
\usepackage{array}
\usepackage{graphicx}
\usepackage{subfig}
\usepackage{enumerate}

\let\OLDthebibliography\thebibliography
\renewcommand\thebibliography[1]{
  \OLDthebibliography{#1}
  \setlength{\parskip}{0pt}
  \setlength{\itemsep}{0pt plus 0.3ex}
}

\begin{document}

\begin{center}
{\LARGE { Worldsheet instantons and coupling selection rules \\ 
\vspace{0.3cm}
 in heterotic orbifolds}}

\renewcommand{\thefootnote}{\alph{footnote}}
\vspace{0.8 cm} {\large Susha L. Parameswaran$^a$ and Ivonne Zavala$^b$
}

\vspace{0.5 cm}
{\small $^a$ Leibniz Universit\"at Hannover, Welfengarten 1, 30167 Hannover, Germany}\\
{\small $^b$ 
University of Groningen,  Nijenborgh 4, 9747 AG Groningen, The Netherlands}



\end{center}

\begin{quotation}
We review recent results \cite{KPZ,CKMPSZ1,CKMPSZ2} on string coupling selection rules for heterotic orbifolds, derived using conformal field theory.  Such rules are the first step towards understanding the viability of the recently obtained compactifications with potentially realistic particle spectra.  They arise from the properties of the worldsheet instantons that mediate the couplings, and include stringy effects that would seem `miraculous' to an effective field theory observer.   
\end{quotation}

\section{Introduction}

The last decade has seen the discovery of a rich landscape of explicit string compactifications with potentially realistic particle spectra.  We now have to understand the dynamics of these constructions, starting with their moduli stabilization, decoupling of exotics, phenomenologically relevant couplings, and eventually addressing cosmology. 
Heterotic orbifold compactifications provide tractable global theories in which the low energy effective field theory (LEEFT) can in principle be computed, and dynamical issues can be addressed, explicitly.  However, it is noteworthy that 25 years after the first string coupling computations 
\cite{HV, Dixon}, we still only have a good understanding of trilinear superpotential 
couplings between massless ground states, moreover, in the absence of the discrete Wilson lines that are necessary for realistic particle spectra.

In this review, we describe recent progress in understanding L-point couplings between massless ground and excited states.  Our focus is on contributions to the holomorphic superpotential, which can be directly inferred from string tree-level correlation functions.  Couplings between twisted states, typically located at different fixed points in the orbifold geometry, are mediated by classical worldsheet instantons stretching between the participating strings.  Properties of the worldsheet instantons lead to several stringy selection rules.  One of these can be interpreted as an R-charge conservation law in the LEEFT.  The others seem to evade a field theoretical interpretation in terms of symmetries, and may be interpreted as `stringy miracles'.  We also discuss anomalies in the R-charge conservation law, and point out an important open question regarding its consistency.

\section{Interactions on orbifolds}
We focus on $\mathbbm{T}^6/\Z_N$ orbifolds, where the generator of the $\Z_N$  point group can be written as $\theta = \textrm{diag}\left(e^{2\pi \imath v_1}, e^{2\pi \imath v_2},e^{2\pi \imath v_3} \right)$, using complex coordinates for the torus.
When compactifying the heterotic string on an orbifold, new twisted states emerge.  For example, the internal space bosonic coordinates for states in the $k$th twisted sector close under the twisted boundary conditions ($i=1,2,3$):
\begin{equation}
 X^i(e^{2\pi {\rm i}}z,e^{-2\pi {\rm i}}\bar{z})=(gX)^i(z,\bar{z})=(\theta^k X)^i(z,\bar{z})+\lambda^i\,
\label{mono}
\end{equation}
with $g=(\theta^k,\lambda)\in S$ and $\lambda^i\in \Lambda$, for $S$ the space group $\Z_N\ltimes\Lambda$ and $\Lambda$ the torus lattice.  Strings closed by $g$ and conjugate elements $\left\{hgh^{-1} |h \in S \right\}$ are physically equivalent.  For orbifolds with non-prime ordered point groups, this implies that physical states may be linear combinations of states localized at different fixed points on the torus: 
\be
|\psi\rangle = |f\rangle + e^{-2\pi\imath \gamma}|\theta f \rangle + \dots + e^{-2\pi\imath (l-1)\gamma}|\theta^{l-1} f \rangle \,, \label{psi}
\ee
with $l$ the smallest integer such that $\theta^{l} f = f + \lambda$ and $\gamma$ the $\gamma$-phase, fixed by the condition that the full heterotic string state be twist invariant.

The vertex operators describing the emission of a twisted field are given, in the limit of vanishing 4D momentum, by
\bea 
  V_{a} = e^{a\phi} \prod_{i=1}^3 (\partial X^i)^{{\cal N}_L^i} \,\,
(\partial \bar X^i)^{\bar{\cal N}_L^i} \,\, e^{\imath q_{sh}^{a\,m} H^{m}}
\, e^{\imath p_{sh}^I X^I} \,
\sigma^{i}_{\left(k,f\right)} \,,  
\eea
where  $a=-1, -1/2$ for spacetime bosonic and fermionic fields respectively.  
Here $X^I$ are the left-moving gauge degrees of freedom (carrying shifted gauge momentum $p_{sh}$), $H^m$ represent the right-moving fermions after bosonization (carrying shifted H-momentum $q^a_{sh}$) and there are a number ${\cal N}_L^i,\, {\bar{\cal N}_L^i}$ of left-moving bosonic oscillator excitations.  The twist fields, $\sigma^{i}_{\left(k,f\right)}$, implement the non-trivial boundary conditions of twisted states, and may be decomposed into a number of auxiliary twist fields in analogy to (\ref{psi}):
\be
\sigma_{(k,\psi)} = \sum_{r=0}^{l-1} e^{-2\pi\imath r \gamma}\sigma_{(k,\theta^r f)} \,. \label{auxsig}
\ee
The scalar $\phi$ is the superconformal ghost field, and physically equivalent vertex operators carrying different ghost-charge can be obtained from the picture-changing operator, e.g.
\bea
V_{0} = \sum_{j=1}^3 \!\left(e^{i q_{0}^{j\,m} H^m} \,
\bar\partial X^j+ e^{-i q_0^{j \,m} H^m} \,\bar\partial \bar X^j \right) 
e^{\phi}\,V_{-1} \,,
\label{V0} 
\eea
which introduces additional H-momentum $q_0^{i} =\delta^i_j$, and right-moving bosonic oscillators.  

We now have all the ingredients to compute the string correlation functions corresponding to terms in the holomorphic superpotential $\left\langle V_{-\frac12} V_{-\frac12} V_{-1}V_{0}^{\L-3}\right \rangle$.  Since the CFT is free, the correlation function factorizes according to the various components of the heterotic string.  The parts corresponding to $H^m$, $X^I$ and ghosts are easily computed and lead to the conservation of gauge momentum (gauge invariance), H-momentum (this fixes the right-moving oscillator numbers according to ${\cal N}_R^i=0$, $\sum_{\alpha} q_{sh \,\alpha}^i - \bar{\cal N}_R^i = 1 \label{Hmomentum}$), as well as ghost-charge. 
The twist field contributions are more complicated, and lead first 
to the so-called space group selection rule on the boundary conditions or space group conjugacy classes, $\prod_{\alpha=1}^\L \left[ g_{\alpha} \right] = (\mathbbm{1},0)$. Imposing these results, the non-trivial part of the correlation function simplifies to:
\be
{\mathcal F} = \prod_{i=1}^3 \left\langle (\partial X^i)^{{\cal N}_L^i}
\,\, (\partial \bar X^i)^{\bar{\cal N}_L^i}  \,\, (\bar\partial
\bar X^i)^{\bar{\cal N}_R^i}  
\sigma_{\left(k_1,\psi_1\right)}^i \cdots
\sigma_{\left(k_\L,\psi_\L\right)}^i 
\right\rangle \, , \label{beforesplit}
\ee
where from now on $\calN_L^i, \bar\calN_L^i$, $\bar\calN_R^i$, refer to the total number of oscillators.   
By using (\ref{auxsig}), ${\mathcal F}$ can be decomposed into a sum of auxiliary correlation functions, weighted by the $\gamma$-phases.

\section{Worldsheet instanton selection rules}
To make further progress, we split the bosonic coordinates into their classical instanton solutions to the equation of motion, $\partial\bar\partial X^i_{cl} = 0$, and their quantum fluctuations:
$X^i(z,\bar z) = X^i_{cl}(z,\bar z) + X^i_{qu}(z,\bar z)$
and in the path integral we sum over all instantons.  
For 3-point couplings ($\alpha=1,2,3$; see \cite{KPZ} for L-point couplings), they are given by:
\bea\label{dXcl}
\partial X^i_{cl}(z) =\prod_{\alpha=1}^3 a^i (z-z_\alpha)^{\k^i_\alpha-1}\,, 
 \qquad \quad
\partial \bar X^i_{cl}(z) = \prod_{\alpha=1}^3\bar b^i (z- z_\alpha)^{-\k^i_\alpha}
\eea
where the functional form was determined by the local monodromy conditions, whereas the constant coefficients, $a^i, b^i$, are determined by the global monodromy conditions to be proportional to torus coset lattice vectors, $\nu^i=(f_2 - f_1 + (1-\theta^{k_1+k_2})(1-\theta^{\gcd(k_1,k_2)})^{-1}\lambda)$.  Also, we have defined $\k_\alpha^i = k_\alpha  \, v^i \mod
1$, such that $0 < \k_\alpha^i \leq 1$ ($0 \leq \k_\alpha^i < 1$) in the first (second)  equation of (\ref{dXcl}). 

Depending on the twisted sectors involved, it may be that the only way to satisfy the local monodromy conditions with a convergent classical action, is with a vanishing solution.  Meanwhile, if the fixed points involved in two instanton solutions are related by the orbifold twist, then so will be the coefficients $a^i, b^i$.  Consequently, symmetries in the orbifold geometry are observed in the set of classical solutions.  We must also consider the quantum part of the correlation function.  There, the basic OPEs of the theory imply that $\langle (\partial X^i_{qu})^s (\partial \bar X^i_{qu})^t  (\bar \partial \bar X^i_{qu})^u \sigma \dots \sigma \rangle = 0$ unless $s=t+u$.  Note also that, as far as quantum properties go, the auxiliary twist fields $\sigma_{(k,\theta^r f)}$ with various $\theta^r f$ are indistinguishable.  By plugging these results into the correlation function and performing some elementary manipulations, we find that the correlation functions vanish unless certain conditions are satisfied.  For factorizable orbifolds we have (see \cite{CKMPSZ1} for non-factorizable orbifolds):

\vspace{0.2cm}
\noindent
{\it Forbidden instanton selection rule (rule 5):}  Applies when twisted sectors are such that there are no non-trivial worldsheet instantons  \cite{KPZ}.  Non-trivial holomorphic (anti-holomorphic) instantons exist if and only if 
  $1 + \sum_\alpha (-1 + \k^i_\alpha) < 0$ ($1 
+
  \sum_\alpha (- \k_\alpha^i) <0$).  If no instantons are allowed, then we require 
${\cal N}_L^i = \bar{\cal N}_L^i + \bar{\cal N}_R^i$.  If only  holomorphic instantons are allowed, then 
${\cal N}_L^i \geq \bar{\cal N}_L^i$.  If instead only anti-holomorphic instantons are allowed, then ${\cal N}_L^i \leq \bar{\cal N}_L^i+\bar{\cal N}_R^i$.

\vspace{0.2cm}
\noindent
{\it R-charge conservation:}  Due to symmetries relating fixed points, generated by $\theta_j$ for planes $j$ with prime ordered twists, and $\prod_{i\neq j}\theta_i$ for non-prime planes $i\neq j$  \cite{CKMPSZ1}.  These imply respectively  $\calN_L^j - \bar\calN_L^j - \bar\calN_R^j = 0 \mod \N^j$ and $\sum_{i\neq j} \N v^i(\calN_L^i - \bar\calN_L^i - \bar\calN_R^i) - \N \sum_\alpha \gamma_\alpha = 0 \mod \N$.

\vspace{0.2cm}
\noindent
{\it Coset vector selection rule (rule 6):}  For couplings with $\K < \N$, for $\K$ the lowest common multiple of the twisted sectors (or their conjugates), instantons enjoy a further symmetry generated by $\theta_i^\K$, leading to $\calN_L^i - \bar\calN_L^i - \bar\calN_R^i = 0 \mod \frac{\N^i}{\K}$ \cite{CKMPSZ2}. 

\vspace{0.2cm}
\noindent
{\it Torus lattice selection rule (rule 4):}  When all twisted sectors are at the same fixed point in plane $i$ for every auxiliary coupling, the symmetry is enhanced from the point group to that of the full torus lattice (order $\M^i$), and we have $\calN_L^i - \bar\calN_L^i - \bar\calN_R^i = 0 \mod \M^i$ \cite{Rule4, KPZ}.

\vspace{0.2cm}
Notice that, because rules 4 and 6 depend on the relative properties of the states that are coupling, they cannot be interpreted in terms of symmetries in the LEEFT.  Rule 5 also appears to be evade field theoretic interpretation; due to the dependence on the right-moving oscillators, it is difficult to assign meaningful, picture-independent charges to the states that are coupling.  Therefore, among the worldsheet instanton selection rules, only the R-charge conservation law  admits an interpretation as a symmetry within the LEEFT.

\section{More on R-charge conservation laws}
Above, we described the derivation of an R-charge conservation law by explicit computation of vanishing correlation functions.  One can also derive R-charge conservation laws by studying directly how the twisted states transform under the remnants of the 10D Lorentz symmetries surviving the orbifold compactification \cite{Nilles:2013lda, CKMPSZ2}.  In this way, we can also study models that include discrete Wilson lines, for whom explicit correlation functions are not yet understood. 
Moreover one should obtain the conservation law directly for the physical couplings, not necessarily superpotential ones.  E.g. consider rotations $\varrho$ of the torus lattice, which leave the fixed-points on the orbifold invariant ($\theta_j$ and $\prod_{i\neq j} \theta_i$ above; see \cite{CKMPSZ2} for more general orbifold isometries leading to new R-symmetries).  By definition, given a $g \in S$, $\varrho (g)$ is conjugate to $g$, and hence there exists a space group element $h_g$, easily found, such that
\begin{equation}
\varrho (g)=h_g g h_g^{-1}\, .
\label{varrho}
\end{equation}  
Using (\ref{varrho}) it is straightforward to
 derive the corresponding R-charges.  The result is
   $\calN_L^j - \bar\calN_L^j - \bar\calN_R^j = 0 \mod \N^j$ for prime planes,  and $\sum_{i\neq j} \N v^i(\calN_L^i - \bar\calN_L^i - \bar\calN_R^i) + \N \sum_\alpha \gamma_\alpha = 0 \mod \N$ for non-prime planes \cite{CKMPSZ2}. Surprisingly, although the origin of both our R-symmetries lies in the orbifold geometry, their R-charges are different.  

We can compute the anomalies for both R-symmetries.  Discrete symmetries are global, but if they arise as remnants of some spontaneously broken gauge symmetry like Lorentz invariance, they should be anomaly free. In heterotic orbifolds, anomalies are generally expected to be universal if they are to be cancelled via the Green-Schwarz mechanism (an exception is discrete target space modular invariance whose anomalies are partially cancelled by threshold corrections).  Beautifully, the R-symmetries derived 
by directly computing R-charges \cite{CKMPSZ2} turn out to have universal anomaly coefficients.  In contrast, the R-symmetries derived by computing explicitly vanishing correlation functions \cite{CKMPSZ1} have non-universal anomalies, suggesting that they might be stringy zeroes that are not observed in the final physical couplings of the full quantum LEEFT.  It remains an essential task to verify the universally anomalous R-symmetry by explicit computation of string couplings, and also to identify the physics behind the discrepancy in the two R-charge conservation laws.  Another important issue is that in order to identify allowed couplings in the final LEEFT, including e.g. effective holomorphic couplings after moduli stabilization, we must also understand the allowed couplings in the K\"ahler potential.

\vskip0.3cm

It is our pleasure to acknowledge and thank our co-authors Nana Cabo-Bizet, Tatsuo Kobayashi, Dami\'an Mayorga-Pe\~na, Sa\'ul Ramos-S\'anchez and Matthias Schmitz.

{\def\section*#1{}

}


\begin{thebibliography}{99}



\bibitem{KPZ}
  Kobayashi T., Parameswaran S., Ramos S., Zavala I. // 
  JHEP 2012. V. 1205. P. 8. [Erratum-ibid. 2012. V. 1212. P. 49.] 

\bibitem{CKMPSZ1}
  Cabo N., Kobayashi T., Mayorga D., Parameswaran S., Schmitz M., Zavala I. // 
  JHEP 2013. V. 1305. P. 76. 

\bibitem{CKMPSZ2}
  Cabo N., Kobayashi T., Mayorga D., Parameswaran S., Schmitz M., Zavala I. // 
  arXiv:1308.5669 [hep-th].
  
\bibitem{HV}
Hamidi S., Vafa C. // 
Nucl.\ Phys.\ B. 1987. V. 279. P. 465. 

\bibitem{Dixon}
Dixon L., Friedan D., Martinec E., Shenker S. // 
Nucl.\ Phys.\ B. 1987. V. 282. P. 13.
  
\bibitem{Rule4}
Font A., Ib\'a\~nez L., Nilles H. P., Quevedo F. // 
Nucl.\ Phys.\  B. 1988. V. 307. P. 109. [Erratum-ibid. 1988. V. 310. P. 764.] 

\bibitem{Nilles:2013lda}
  Nilles H. P., Ramos S., Ratz M., Vaudrevange P. // 
  Phys.\ Lett.\ B. 2013. V. 726. P. 876. 
    
\end{thebibliography}
\end{document}